\documentclass[aps,prd,nofootinbib,preprint]{revtex4}

\usepackage{graphicx}
\usepackage{psfrag}

\begin{document}
\def\dirac#1{#1\llap{/}}
\def\pv#1{\mathbf{#1}_\perp}

\title{$B\to K$ Transition Form Factor up to ${\cal O}(1/m^2_b)$ within the $k_T$ Factorization Approach}
\author{Xing-Gang Wu$^{1}$ \footnote{email: wuxg@cqu.edu.cn},
Tao Huang$^{2}$\footnote{email: huangtao@mail.ihep.ac.cn} and
Zhen-Yun Fang$^{1}$ \footnote{email: zyfang@cqu.edu.cn}}
\address{$^1$Department of Physics, Chongqing University, Chongqing 400044,
P.R. China\\ $^2$Institute of High Energy Physics, Chinese Academy
of Sciences, P.O.Box 918(4), Beijing 100049, P.R. China}

\begin{abstract}
In the paper, we apply the $k_T$ factorization approach to deal with
the $B\to K$ transition form factor $F^{B\to K}_{+,0}(q^2)$ in the
large recoil regions. The B-meson wave functions $\Psi_B$ and
$\bar\Psi_B$ that include the three-particle Fock states'
contributions are adopted to give a consistent PQCD analysis of the
form factor up to ${\cal O} (1/m^2_b)$. It has been found that both
the wave functions $\Psi_B$ and $\bar\Psi_B$ can give sizable
contributions to the form factor and should be kept for a better
understanding of the $B$ meson decays. Then the contributions from
different twist structures of the kaon wavefunction are discussed,
including the $SU_f(3)$-breaking effects. A sizable contribution
from the twist-3 wave function $\Psi_p$ is found, whose model
dependence is discussed by taking two group of parameters that are
determined by different distribution amplitude moments obtained in
the literature. It is also shown that $F^{B\to
K}_{+,0}(0)=0.30\pm0.04$ and $[F^{B\to K}_{+,0}(0)/F^{B\to
\pi}_{+,0}(0)]=1.13\pm0.02$, which are more reasonable and
consistent with the light-cone sum rule results in the large recoil
regions. \\

\noindent {\bf PACS numbers:} 12.38.Aw, 12.38.Bx, 13.20.He, 14.40.Aq

\end{abstract}

\maketitle

\section{Introduction}

A study on the heavy-to-light exclusive processes plays a
complementary role in determination of the fundamental parameters of
the standard model and in developing the QCD theory. And there is an
increasing demand for more reliable QCD calculations of the
heavy-to-light form factors. We have done a consistent analysis of
the $B\to\pi$ transition form factor in Refs.\cite{hwbpi,hqw}, which
shows that the results from the PQCD approach, the lattice QCD
approach and the QCD light cone sum rules (LCSRs) are complementary
to each other and by combining the results of those three
approaches, one can obtain an understanding of the $B\to\pi$
transition form factor in the whole physical regions. It is argued
that by applying the $k_T$ factorization approach
\cite{lbhm,lb,lis}, where the transverse momentum dependence for
both the hard scattering part and the non-perturbative wavefunction,
the Sudakov effects and the threshold effects are included, one can
regulate the endpoint singularity from the hard scattering part
effectively and derive a more reliable PQCD result of the B-meson
decays. Furthermore, by applying the B-meson wave functions up to
next-to-leading Fock state, Ref.\cite{hqw} calculated the $B\to\pi$
transition form factor up to ${\cal O}(1/m_b^2)$ and also made some
discussions on the reasonable regions for the two phenomenological
B-meson wave function parameters $\bar\Lambda$ and $\delta$, where
$\bar\Lambda$ is the effective mass of B meson that determines the
B-meson's leading Fock state behavior and $\delta$ is a typical
parameter that determines the broadness of the B-meson transverse
distribution. Since both pion and kaon are pseudo-scalar mesons, it
will be interesting to give a consistent PQCD analysis of the $B\to
K$ transition form factor up to order ${\cal O}(1/m^2_b)$ based on
the results of $B\to\pi$ transition form factor.

In the literature, the $B\to K$ transition form factor has been
studied under several approaches
\cite{lucai,lucai2,huangli,sumrule,pballsum1,nlosum}. A PQCD
calculation has been done in Ref.\cite{lucai}, which can be roughly
treated as a leading-order estimation ${\cal O}(1/m_b)$ since some
of the power suppressed terms both in the hard scattering amplitude
and the B-meson wave function have been neglected. The $B\to K$
transition form factor has also been analyzed by several groups
under the QCD LCSR approach \cite{huangli,sumrule,pballsum1,nlosum},
where some extra treatments to the correlation function either from
the B-meson side or from the kaonic side are adopted to improve
their LCSR estimations. New sum rule for the $B\to K$ form factor is
derived by expanding the correlation function near the light cone in
terms of B-meson distributions \cite{nlosum}, in which the
contributions of the quark-antiquark and quark-antiquark-gluon
components in the B-meson are taken into account. While in
Ref.\cite{huangli} an improved LCSR approach that had been raised in
Ref.\cite{huangsr1} was adopted to eliminate the contributions from
the most uncertain kaonic twist-3 wave functions and to enhance the
reliability of sum rule calculations of the $B_s\to K$ form factor.
A systematic QCD LCSR calculation has been done in
Ref.\cite{sumrule} by including the one-loop radiative corrections
to the kaonic twist-2 and twist-3 contributions, and the kaonic
leading-order twist-4 corrections. It can be found that the main
uncertainties in estimation of the $B\to K$ transition form factor
come from the B-meson wave function and the kaonic twist-2 and
twist-3 wave functions.

In doing the PQCD calculations on the B-meson decays, an important
issue is whether we need to take both the two wavefunctions $\Psi_B$
and $\bar\Psi_B$ into consideration or simply $\Psi_B$ is enough? By
taking the frequently used first type definition of
$\Psi_B=\frac{\Psi_B^{+}+\Psi_B^-}{2}$ and
$\bar{\Psi}_B=\frac{\Psi_B^{+}-\Psi_B^-}{2}$, where $\Psi_B^{\pm}$
are defined in Ref.\cite{hqet}, it can be found that
\cite{wu1,descotes} their distribution amplitudes have quite
different endpoint behaviors even under the Wandzura-Wilczek (WW)
approximation \cite{wwlike}, such difference may be strongly
enhanced by the hard scattering kernel. For example, the ratio
between the contributions of $\bar\Psi_B$ and $\Psi_B$ is about
$(-70\%)$ \cite{weiy2,hwbpi} for the $B\to\pi$ from factor in the
large recoil regions. So the contribution from $\bar\Psi_B$ under
the above definition can not be neglected and it is needed to
suppress the big contribution coming from $\Psi_B$ so as to obtain a
reasonable total contributions. To derive more accurate estimation,
Ref.\cite{lucai} raised the second type definition of
$\Psi_B=\Psi_B^{+}$ and $\bar{\Psi}_B=(\Psi_B^{+}-\Psi_B^-)$, under
which the contribution from $\bar\Psi_B$ is of order ${\cal O}
(1/m_b)$ to that of $\Psi_B$ \cite{li1}. For convenience, in the
following, we shall adopt the second type definition of $\Psi_B$ and
$\bar\Psi_B$ to do our calculation. Then one may ask is it enough to
give a ${\cal O}(1/m^2_b)$ estimation with $\Psi_B$ and $\bar\Psi_B$
under the WW approximation ? As has been pointed out in
Ref.\cite{charng}, the 3-particle Fock states' contributions to the
B-meson wave function can be estimated by attaching an extra gluon
to the internal off-shell quark line, and then $(1/m_b)$ power
suppression is induced in comparison to that of the WW-part B-meson
wave functions. Recently, the B-meson light-cone wave functions have
been investigated up to next-to-leading order in Fock state
expansion in the heavy quark limit \cite{hqw}. It was shown that by
using the relations between 2- and 3- particle wave functions
derived from the QCD equations of motion and the heavy-quark
symmetry, one can give a constraint on the transverse momentum
dependence of the B-meson wave function, whose distribution tends to
be a hyperbola-like curve other than a simple delta function that is
derived under the WW approximation. These results provide us a
chance to give a consistent PQCD analysis of the $B\to K$ form
factor up to order ${\cal O}(1/m^2_b)$.

Another issue we need to be more careful is about the kaonic wave
functions. The distribution amplitude (DA) for the twist-2 wave
function $\Psi_K$ has been deeply studied, e.g. by the light-front
quark model \cite{quark1}, the LCSR approach
\cite{lcsr1,pballa1k,braunlenz,ballmoments} and the lattice
calculation \cite{lattice1,lattice2} and etc. In Ref.\cite{lcsr1},
the QCD sum rule for the diagonal correlation function of local and
nonlocal axial-vector currents is used, in which the contributions
of condensates up to dimension six and the ${\cal
O}(\alpha_s)$-corrections to the quark-condensate term are taken
into account. The first Gegenbauer moment $a^K_1(1{\rm GeV})$ of the
twist-2 DA derived there, i.e. $a^K_1(1{\rm GeV})= 0.05\pm 0.02$, is
consistent with that of the lattice calculations
\cite{lattice1,lattice2}, so we shall constrain $a^K_1(1{\rm GeV})$
within this range when constructing a model for $\Psi_K$. As for the
twist-3 wave function $\Psi_p$, the calculations of it has more
uncertainty than that for the leading twist, e.g. its DA moments in
Refs.\cite{ballmoments,hzw,instaton} are quite different from each
other, where the DA moments in Refs.\cite{ballmoments,hzw} are
derived by using the QCD light-cone sum rules and the moments in
Ref.\cite{instaton} are derived based on the effective chiral action
from the instanton. Under the PQCD approach, according to our
experience on the $B\to\pi$ transition form factor \cite{hwbpi} and
the pion electro-magnetic form factor \cite{pitwist3}, it can be
found that for a twist-3 wave function with a better endpoint
behavior other than the asymptotic one, the twist-3's contributions
are indeed power suppressed to the leading twist's contribution that
favor the conventional power counting rules. In the present paper,
we shall adopt two groups of DA moments \cite{ballmoments,instaton}
together with the Brodsky-Huang-Lepage (BHL) prescription~\cite{bhl}
to construct a model for $\Psi_p$, and then make a discussion on its
uncertainty to the $B\to K$ transition form factor. The
$SU_f(3)$-breaking effects shall also be included for constructing
the kaonic wave functions.

The purpose of the paper is to reexamine the $B\to K$ transition
form factor in the PQCD $k_T$ factorization approach up to ${\cal
O}(1/m^2_b)$. Under the $k_T$ factorization approach, the full
transverse momentum dependence ($k_T$-dependence) for both the hard
scattering part and the non-perturbative wave function, the Sudakov
effects and the threshold effects are included to cure the endpoint
singularity. Furthermore, we shall analyze the power suppressed
contributions from both the wave functions and the hard scattering
amplitude and then give a consistent analysis of the form factor up
to ${\cal O}(1/m^2_b)$, which have not been considered in the
literature. In section II, we give the calculated technology for the
form factor in the large recoil regions. Also we present the model
wave functions of the kaon with better endpoint behavior in the same
section, which are constructed based on the BHL
prescription~\cite{bhl} and the DA moments obtained in
Ref.\cite{ballmoments,instaton}. In section III, we give our
numerical results. Conclusion and a brief summary are presented in
the final section.

\section{calculation technology for the $B\to K$ transition form factor}

The $B\to K$ transition form factors $F_+^{B\to K}(q^2)$ and
$F_0^{B\to K}(q^2)$ are defined as follows:
\begin{equation}\label{eq:bk1}
\langle K(P_K)|\bar s \gamma_{\mu}b|\bar
B(P_B)\rangle=\left[(P_B+P_K)_{\mu}-
\frac{M_B^2-M_K^2}{q^2}q_{\mu}\right] F_+^{B\to K}(q^2)+
 \frac{M_B^2-M_K^2}{q^2} q_{\mu}F_0^{B\to K}(q^2)
\end{equation}
where $F_+^{B\to K}(0)$ should be equal to $F_0^{B\to K}(0)$ so as
to cancel the poles at $q^2=0$. The amplitude for the $B\to K$
transition form factor can be factorized into the convolution of the
wave functions for the respective hadrons with the hard-scattering
amplitude. In the large recoil regions, the $B\to K$ transition form
factor is dominated by a single gluon exchange in the lowest order.
In Ref.\cite{hwbpi}, we have done a consistent analysis of the
$B\to\pi$ transition form factor within the $k_T$ factorization
approach, where the power suppressed terms up to ${\cal O}(1/m^2_b)$
have been kept explicitly in the hard scattering amplitude. The
interesting reader may refer to Ref.\cite{hwbpi} for more details
\footnote{Three typo errors are found in Ref.\cite{hwbpi}, i.e. in
Eq.(3) $\frac{\slash\!\!\! P_B\, M_B}{2}$ should be changed to
$\frac{\slash\!\!\! P_B+M_B}{2}$, in Eq.(5) the factor
$[3-\eta-x\eta]$ should be changed to $[3-\eta+x\eta]$ and in Eq.(7)
$y$ should be changed to $\eta$.}. More specifically, for the
present case, one needs to know the momentum projection for the
matrix element of the kaon and B meson in deriving the hard
scattering amplitude. By keeping the transverse momentum dependence
in the wave function, the momentum projection for the matrix element
of the kaon has the following form,
\begin{equation}
M_{\alpha\beta}^{K} = \frac{i f_{\pi}}{4} \Bigg\{
\slash\!\!\!p\,\gamma_5\,\Psi_{K}(x, \mathbf{k_\perp})-
\mu_K\gamma_5 \left(\Psi_p(x, \mathbf{k_\perp})
-i\sigma_{\mu\nu}\left(n^{\mu}\bar{n}^{\nu}\,\frac{\Psi_{\sigma}'(x,
\mathbf{k_\perp})}{6}-p^\mu\,\frac{\Psi_\sigma(x,
\mathbf{k_\perp})}{6}\, \frac{\partial}{\partial
\mathbf{k}_{\perp\nu}} \right)\right)
\Bigg\}_{\alpha\beta},\label{benek}
\end{equation}
where $f_{K}$ is the kaon decay constant and $\mu_K$ is the
phenominological parameter $\mu_K=M_K^2/(m_s+m_u)$, which is a scale
characterized by the chiral perturbation theory.
$\Psi_{K}(x,\mathbf{k_{\perp}})$ is the twist-2 wave function,
$\Psi_p(x,\mathbf{k_{\perp}})$ and
$\Psi_{\sigma}(x,\mathbf{k_{\perp}})$ are twist-3 wave functions,
respectively. $\Psi_{\sigma}'(x, \mathbf{k_\perp})=\partial
\Psi_{\sigma} (x, \mathbf{k_\perp})/\partial x$,
$n=(\sqrt{2},0,\mathbf{0}_\bot)$ and
$\bar{n}=(0,\sqrt{2},\mathbf{0}_\bot)$ are two null vectors that
point to the plus and the minus directions, respectively. And the
momentum projection for the matrix element of the B meson can be
written as \cite{weiy2,BenekeFeldmann}:
\begin{equation}\label{projectorB}
M^B_{\alpha\beta}=-\frac{if_B}{4}
\left\{\frac{\dirac{p}_B+M_B}{2}\left[\dirac{n}
\Psi^+_B(\xi,\mathbf{l_\bot}) +
\bar\dirac{n}\Psi^-_B(\xi,\mathbf{l_\bot})-\Delta(\xi,
\mathbf{l}_{\bot}) \gamma^\mu \frac{\partial} {\partial
l_\perp^\mu}\right] \gamma_5\right\}_{\alpha\beta}\ ,
\end{equation}
where $\xi=\frac{l^+}{M_B}$ is the momentum fraction for the light
spectator quark in the B meson and $\Delta(\xi, \mathbf{l}_{\bot})
=M_B \int_0^{\xi} d\xi' (\Psi^-_B(\xi',\mathbf{l}_{\bot})
-\Psi^+_B(\xi',\mathbf{l}_{\bot}))$. The four-component
$l_\perp^\mu$ in Eq.(\ref{projectorB}) is defined through,
$l^{\mu}_\perp=l^\mu-\frac{(l^+ n^\mu +l^-\bar{n}^\mu)}{2}$ with
$l=(\frac{l^+}{\sqrt{2}},\frac{l^-}{\sqrt{2}},\mathbf{l}_\perp)$. By
including the Sudakov form factors and the threshold resummation
effects, one can obtain the formulae for the $B\to K$ transition
form factors $F_+^{B\to K}(q^2)$ and $F_0^{B\to K}(q^2)$ in the
transverse configuration $b$-space, which can be simply obtained
from Ref.\cite{hwbpi} by changing the pion wave functions to the
present case of kaon and by changing $\Psi_B$ and $\bar\Psi_B$ to
the second type definition as described in the INTRODUCTION.

In PQCD approach, the parton transverse momenta $\mathbf{k}_\perp$
are not negligible around the endpoint region. The relevant Sudakov
factors from both $k_\perp$ and the threshold resummation
\cite{threshold1} can cure the endpoint singularity which makes the
calculation of the hard amplitudes infrared safe, and then the main
contribution comes from the perturbative region. Also it is
necessary to keep the transverse momentum dependence in the wave
functions to derive a more reliable estimation in PQCD. In
principle, the Bethe-Salpeter formalism~\cite{bs} and the
discretized light cone quantization approach~\cite{dlcq} could
determine the hadronic wave functions, but in practice there are
many difficulties in getting the exact wave functions at present.
The BHL prescription \cite{bhl}, which connects the equal-time wave
function in the rest frame and the wave function in the infinite
momentum frame, provides a useful way to use the approximate bound
state solution of a hadron in terms of the quark model as the
starting point for modeling the hadronic wave function. So in the
present paper, we will adopt the BHL prescription for constructing
the kaonic wave functions. While for the $B$-meson wave function,
they have been investigated up to next-to-leading order in Fock
state expansion in the heavy quark limit in Ref.\cite{hqw}, which
shall be adopted to do our discussions.

A simple model has been raised in Ref.\cite{hqw} for the B-meson
wave functions $\Psi^{+}_{B}$ and $\Psi^{-}_{B}$, which keep the
main features caused by the 3-particle Fock states and whose
transverse momentum dependence are still the like-function of the
off-shell energy of the valence quarks but shall broaden the
transverse momentum dependence under the WW approximation to a
certain degree. And in the compact parameter $b_B$-space, it reads
\cite{hqw}
\begin{equation}\label{newmodel1}
\Psi^{+}_{B}(\xi,b_B)=(16\pi^3)\frac{M_B^2\xi}{\omega_0^2}\exp
\left( -\frac{M_B\xi}{\omega_0}\right) \Big(\Gamma[\delta]
J_{\delta-1} [\kappa] +(1-\delta)\Gamma[2-\delta]
J_{1-\delta}[\kappa]\Big)\left( \frac{\kappa}{2} \right)^{1-\delta}
\end{equation}
and
\begin{equation}
\Psi^{-}_{B}(\xi,b_B)=(16\pi^3)\frac{M_B}{\omega_0}\exp \left(
-\frac{M_B \xi}{\omega_0}\right)\Big(\Gamma[\delta] J_{\delta-1}
[\kappa] +(1-\delta)\Gamma[2-\delta] J_{1-\delta}[\kappa]\Big)\left(
\frac{\kappa}{2} \right)^{1-\delta} ,\label{newmodel2}
\end{equation}
where $\omega_0=2\bar\Lambda/3$, $\bar\xi=\bar\Lambda/M_B$ and
$\kappa=\theta(2\bar\xi-\xi) \sqrt{\xi (2\bar\xi-\xi)} M_B b_B $.
The factor $(16\pi^3)$ is introduced to ensure their Fourier
transformation, i.e. $\Psi^{\pm}_{B}(\xi,\mathbf{k}_\perp)$, satisfy
the normalization, $\int \frac{d\xi d^{2}{\bf k}_{\perp}}{16\pi^3}
\Psi^{\pm}_{B}(\xi,\mathbf{k}_\perp) =1$. It can be found that both
$\Psi^{+}_{B}$ and $\Psi^{-}_{B}$ have the same transverse momentum
dependence and only two phenomenological parameters $\bar\Lambda$
and $\delta$ are introduced. $\bar\Lambda$ is the effective mass of
B meson, $\bar\Lambda=M_B-m_b$, which determines the B meson's
leading Fock state behavior. $\delta$ is a typical parameter that
determines the broadness of the B-meson transverse distribution in
comparison to the WW-like one. The WW-like B-meson wave functions in
the compact parameter $b_B$-space can be found in Ref.\cite{hwbpi}.
And a direct comparison shows that when $\delta\to 1$, the
transverse momentum dependence of the B-meson wave function in
Eqs.(\ref{newmodel1},\ref{newmodel2}) returns to a simple
$\delta$-function, which is the same as that of the B-meson wave
function under the Wandzura-Wilczek approximation \cite{wu1,wwapp}.
According to the definitions, we have
$\Psi_{B}(\xi,b_B)=\Psi^{+}_{B}(\xi,b_B)$,
$\bar\Psi_{B}(\xi,b_B)=\Psi^{+}_{B}(\xi,b_B)-\Psi^{-}_{B}(\xi,b_B)$
and $\Delta(\xi,b_B)=-M_B\int^\xi_0 d\xi' \bar\Psi_B(\xi',b_B)$.

Next, we construct the kaonic twist-2 wave function based on its
first Gegenbauer moment $a_1^K$ and on the BHL prescription
\cite{bhl}. The first Gegenbauer moment $a_1^K$ has been studied by
the light-front quark model \cite{quark1}, the LCSR approach
\cite{lcsr1,pballa1k,ballmoments} and the lattice calculation
\cite{lattice1,lattice2} and etc. In Ref.\cite{lcsr1}, the QCD sum
rule for the diagonal correlation function of local and nonlocal
axial-vector currents is used, in which the contributions of
condensates up to dimension six and the ${\cal
O}(\alpha_s)$-corrections to the quark-condensate term are taken
into account. The moments derived there are close to that of the
lattice calculation \cite{lattice1,lattice2}, so we shall take
$a^K_1(1{\rm GeV})=0.05\pm 0.02$ to determine the model wave
function $\Psi_K$. Based on the BHL prescription, we take the
twist-2 wave function of kaon as
\begin{eqnarray}
\Psi_{K}(x,\mathbf{k}_\perp) &=& [1+B_K C^{3/2}_1(2x-1)]\times
\frac{A_K}{x(1-x)} \exp \left[-\beta_K^2
\left(\frac{k_\perp^2+m_q^2}{x}+ \frac{k_\perp^2+m_s^2}
{1-x}\right)\right],\label{wavek}
\end{eqnarray}
where $q=u,\; d$, $C^{3/2}_1(1-2x)$ is the Gegenbauer polynomial. In
comparison to the pion wave function (e.g. \cite{hwpi}), it can be
found that the $SU_f(3)$ symmetry is broken by a non-zero $B_K$ and
by the mass difference between the $s$ quark and $u$ (or $d$) quark
in the exponential factor. The $SU_f(3)$ symmetry breaking in the
lepton decays of heavy pseudoscalar mesons and in the semileptonic
decays of mesons have been studied in Ref.\cite{khlopov}. For
definiteness, we take the conventional values for the constitute
quark masses: $m_q=0.30{\rm GeV}$ and $m_s=0.45{\rm GeV}$. The
parameters $A_K$, $B_K$ and $\beta_K$ can be determined by the value
of $a^K_1$ together with the normalization condition:
\begin{equation}\label{normalization}
\int^1_0 dx \int_{k_\perp^2<\mu_0^2} \frac{d^{2}{\bf
k}_{\perp}}{16\pi^3}\Psi_K(x,{\bf k}_{\perp}) =1
\end{equation}
and the constraint $\langle \mathbf{k}_\perp^2 \rangle^{1/2}_K
\approx \langle \mathbf{k}_\perp^2 \rangle^{1/2}_\pi=0.350{\rm GeV}$
\cite{gh}, where the average value of the transverse momentum square
is defined as
\begin{displaymath}
\langle \mathbf{k}_\perp^2 \rangle^{1/2}_K=\frac{\int dx
d^2\mathbf{k}_\perp |\mathbf{k}_\perp^2| |\Psi_K(x,{\bf
k}_{\perp})|^2} {\int dx d^2\mathbf{k}_\perp |\Psi_K(x,{\bf
k}_{\perp})|^2} .
\end{displaymath}
The parameter $\mu_0$ in the model wave function stands for some
hadronic scale that is of order ${\cal O}(1~{\rm GeV})$. For
clarity, we set $\mu_0=1$ GeV. The DA $\phi_K(x,\mu_0)$ is defined
as $\phi_K(x,\mu_0)=\int_{k_\perp^2<\mu_0^2} \frac{d^{2}{\bf
k}_{\perp}}{16\pi^3} \Psi_K(x,{\bf k}_{\perp})$. The first
Gegenbauer moment $a^K_1(\mu_0)$ of
Refs.\cite{lcsr1,pballa1k,ballmoments} can be defined as
\begin{equation}
a^K_1(\mu_0)=\frac{\int_0^1 dx \phi_K(1-x,\mu_0)C^{3/2}_1(2x-1)}
{\int_0^1 dx 6x(1-x) [C^{3/2}_1(2x-1)]^2} \; ,
\end{equation}
where $\phi_K(1-x,\mu_0)$ other than $\phi_K(x,\mu_0)$ should be
adopted, since in Refs.\cite{lcsr1,pballa1k,ballmoments} $x$ stands
for the momentum fraction of $s$-quark in the kaon ($\bar K$), while
in the present paper we take $x$ as the momentum fraction of the
light $q$-(anti)quark in the kaon ($K$) \footnote{In the literature,
there are some ambiguities in use of $\phi_{K,p}(x,\mu_0)$ or
$\phi_{K,p}(1-x,\mu_0)$ in connection to the hard scattering part.
This will cause errors when the $SU_f(3)$-symmetry is broken.}.
Based on the above discussions, we can obtain the values for $A_K$,
$B_K$ and $\beta_K$:
\begin{equation} \label{psika1}
A_K \cong 2.71\times 10^{2} {\rm GeV}^{-1}\;,\; B_K \cong [0.116-0.9
a^K_1(\mu_0)]\;,\; \beta_K \cong 0.877{\rm GeV}^{-1} ,
\end{equation}
where the values of $A_K$ and $\beta_K$ are almost constant, i.e.
their changes $(\delta A_K / A_K)$ and $(\delta\beta_K / \beta_K)$
are less than $0.001$ by varying $a^K_1(\mu_0)$ within the range of
$[0.03,0.07]$. More specifically, for the case of
$a^K_1(\mu_0)=0.05$, we have
\begin{displaymath}
A_K=2.71\times 10^{2} {\rm GeV}^{-1}\;,\;\; B_K=0.071\;,\;\;
\beta_K=0.877{\rm GeV}^{-1}.
\end{displaymath}

As will be seen that the contributions from twist-3 wave function
$\Psi_\sigma(x,\vec{k_\perp})$ is less important in comparison to
that of $\Psi_K(x,\vec{k_\perp})$ and $\Psi_p(x,\vec{k_\perp})$,
which is similar to the case of $B\to\pi$ transition from factor
\cite{hwbpi}. So basing on the BHL prescription, we directly take
the twist-3 wave function $\Psi_{\sigma}$ of kaon as
\begin{eqnarray}
\Psi_{\sigma}(x,\mathbf{k}_\perp) &=& A_\sigma \exp \left[-\beta_K^2
\left(\frac{k_\perp^2+m_q^2}{x}+ \frac{k_\perp^2+m_s^2}
{1-x}\right)\right],\label{waves}
\end{eqnarray}
where $A_\sigma$ can be determined by its normalization condition,
i.e. $A_\sigma=1.36\times10^{3} {\rm GeV}^{-1}$.

As for the twist-3 wave function $\Psi_p(x,\vec{k_\perp})$, its DA's
asymptotic behavior, $\phi^{as}_p(x,\infty)= 1$, so its endpoint
singularity is much more serious. Then the transverse momentum
dependence of $\Psi_p(x,\vec{k_\perp})$ is much more important than
that of $\Psi_K(x,\vec{k_\perp})$ and $\Psi_\sigma(x,\vec{k_\perp})$
in order to cure the endpoint singularity. One can construct
$\Psi_p(x,\vec{k_\perp})$ in the following form,
\begin{eqnarray}\label{wavep}
\Psi_p(x,\vec{k_\perp}) &=& [1+B_p
C^{1/2}_1(2x-1)+C_p C^{1/2}_2(2x-1)]\times\nonumber\\
&&\frac{A_p}{x(1-x)} \exp \left[-\beta_K^2
\left(\frac{k_\perp^2+m_q^2}{x}+ \frac{k_\perp^2+m_s^2}
{1-x}\right)\right],
\end{eqnarray}
where $x$ stands for the light quark $q$'s momentum fraction,
$C^{1/2}_1(2x-1)$ and $C^{1/2}_2(2x-1)$ are Gegenbauer polynomials
and the coefficients $A_p$, $B_p$ and $C_p$ can be determined by its
DA moments. The DA $\phi_p(x,\mu_0)$ is defined as
$\phi_p(x,\mu_0)=\int_{k_\perp^2<\mu_0^2} \frac{d^{2}{\bf
k}_{\perp}}{16\pi^3} \Psi_p(x,{\bf k}_{\perp})$.

\begin{figure}
\centering
\includegraphics[width=0.5\textwidth]{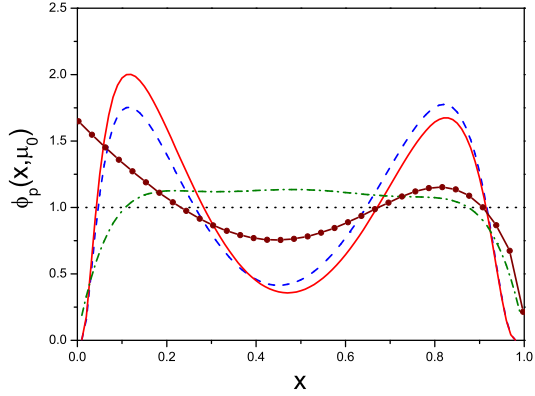}
\caption{Kaon $\phi_p(x,\mu_0)$ with its parameters determined by
the two groups of DA moments \cite{ballmoments,instaton}. The solid
line and the dashed line are for $\phi^1_p(x,\mu_0)$ and
$\phi^2_p(x,\mu_0)$ respectively. For comparison, the big dotted
line and the dash-dot line are for $\phi^{sr}_p(x,\mu_0)$
\cite{ballmoments} and $\phi^{in}_p(x,\mu_0)$ \cite{instaton}
respectively. The dotted line is the asymptotic behavior of
$\phi^{as}_p(x,\infty)= 1$.} \label{DAphip}
\end{figure}

To discuss the uncertainty caused by $\Psi_p$, we take two groups of
DA moments that have been obtained in
Refs.\cite{ballmoments,instaton} to determine the coefficients
$A_p$, $B_p$ and $C_p$, where the moments in Ref.\cite{ballmoments}
are derived by using the QCD light-cone sum rules and the moments in
Ref.\cite{instaton} are derived based on the effective chiral action
from the instanton:
\begin{eqnarray}
{\rm Group \;\;1\;\;\;\;}[23] : && \langle
x^0\rangle^K_p=1,\;\;\langle x^1\rangle^K_p=0.06124,\;\; \langle
x^2\rangle^K_p=0.36757 , \\
{\rm Group \;\;2\;\;\;\;}[27] : && \langle
x^0\rangle^K_p=1,\;\;\langle x^1\rangle^K_p=0.00678,\;\; \langle
x^2\rangle^K_p=0.35162.
\end{eqnarray}
Here the moments are defined as $\langle x^i\rangle^K_p =\int^1_{0}
dx (2x-1)^i\phi_p(1-x,\mu_0)$ with $i=(0,1,2)$. It should be noted
that the moments defined in Ref.\cite{ballmoments,instaton} are for
$\phi_p(1-x,\mu_0)$ other than $\phi_p(x,\mu_0)$, since in these
references $x$ stands for the momentum fraction of $s$-quark in the
kaon ($\bar{K}$), while in the present paper $x$ stands for the
momentum fraction of the light quark $q$ in the kaon ($K$). Taking
the above two groups of DA moments for $\phi_p$, the parameters of
$\Psi_p(x,\vec{k_\perp})$ can be determined as,
\begin{eqnarray}\label{group1}
{\rm Group \;\;1:\;\;}&&\;\;\;A^1_p=382. {\rm GeV}^{-1},\;\;\;
B^1_p=0.311,\;\;\; C^1_p=1.61 ,\\
\label{group2} {\rm Group \;\;2:\;\;}&&\;\;\;A^2_p=422. {\rm
GeV}^{-1},\;\;\; B^2_p=0.257,\;\;\; C^2_p=1.52 .
\end{eqnarray}
The distribution amplitudes for these two group of parameters are
shown in Fig.(\ref{DAphip}), where $\phi^1_p(x,\mu_0)$ is determined
by Group 1 parameters and $\phi^2_p(x,\mu_0)$ is determined by Group
2 parameters respectively. For comparison, we also draw the
distributions derived in Ref.\cite{ballmoments,instaton} in
Fig.(\ref{DAphip}), i.e. $\phi^{sr}_p(x,\mu_0)$ stands for the DA
obtained in Ref.\cite{ballmoments} and $\phi^{in}_p(x,\mu_0)$ stands
for that of Ref.\cite{instaton}. One may observe that different from
$\phi^{sr}_p(x,\mu_0)$ and $\phi^{in}_p(x,\mu_0)$, both
$\phi^1_p(x,\mu_0)$ and $\phi^2_p(x,\mu_0)$ are double humped curves
and are highly suppressed in the endpoint region. Such feature is
necessary to suppress the endpoint singularity coming from the
hard-scattering kernel and then to derive a more reasonable results
for the twist-3 contributions to the $B\to K$ form factor.

It is more convenient to transform the kaon wave functions into the
compact parameter $b_K$-space, which can be done with the help of
the Fourier transformation
\begin{displaymath}
\Psi(x,b_K)=\int_{|\mathbf{\mathbf{k}}|<1/b_K}
d^2\mathbf{k}_\perp\exp\left(-i\mathbf{k}_\perp
\cdot\mathbf{b}_K\right)\Psi(x,\mathbf{k}_\perp),
\end{displaymath}
where $\Psi$ stands for $\Psi_K$, $\Psi_p$ and $\Psi_\sigma$,
respectively. The upper edge of the integration
$|\mathbf{k}_\perp|<1/b_K$ is necessary to ensure that the wave
function is soft enough \cite{huang2}. After doing the Fourier
transformation, we obtain the kaonic wave functions in the compact
parameter $b_K$-space:
\begin{eqnarray}
\Psi_{K}(x,b_K)&=& \frac{2\pi A_K}{x(1-x)}[1+B_K C^{3/2}_1(2x-1)]
\exp \left[-\beta_K^2\left(\frac{m_s^2}{1-x}+
\frac{m_q^2} {x}\right) \right] \nonumber\\
&& \times\int_0^{1/b_K} \exp\left(\frac{-\beta_K^2 k_\perp^2}
{x(1-x)}\right)J_0(b_K k_\perp) k_\perp dk_\perp \;,\nonumber\\
\Psi_{\sigma}(x,b_K)&=&2\pi A_{\sigma} \exp
\left[-\beta_K^2\left(\frac{m_s^2}{1-x}+ \frac{m_q^2} {x}\right)
\right] \int_0^{1/b_K} \exp\left(\frac{-\beta_K^2 k_\perp^2}
{x(1-x)}\right)J_0(b_K k_\perp) k_\perp dk_\perp\nonumber
\end{eqnarray}
and
\begin{eqnarray}
\Psi_p(x,b_K)&=&\frac{2\pi A_p}{x(1-x)}[1+B_p C^{1/2}_1(2x-1)+C_p
C^{1/2}_2(2x-1)] \exp \left[-\beta_K^2\left(\frac{m_s^2}{1-x}+
\frac{m_q^2} {x}\right) \right]\nonumber\\
&& \times \int_0^{1/b_K} \exp\left(\frac{-\beta_K^2 k_\perp^2}
{x(1-x)}\right)J_0(b_K k_\perp) k_\perp dk_\perp \;. \nonumber
\end{eqnarray}

\section{numerical calculations}

In the numerical calculations, we adopt
\begin{displaymath}
\Lambda^{(n_f=4)}_{\over{MS}}=250{\rm MeV},\;  f_B=190{\rm MeV},\;
M_B=5.279{\rm GeV} ,\; f_K=160{\rm MeV},\; M_K=494{\rm MeV}.
\end{displaymath}
As for the phenominological parameter $\mu_K=M_K^2/(m_s+m_u)$, which
is a scale characterized by the chiral perturbation theory, we take
its value to be $\mu_K \simeq 1.70$ GeV.

In the following, we first discuss the properties of $F_+^{B\to
K}(q^2)$ and $F_0^{B\to K}(q^2)$ that are calculated up to ${\cal
O}(1/m^2_b)$, i.e. to show how $F_+^{B\to K}(q^2)$ and $F_0^{B\to
K}(q^2)$ are affected by the B-meson wave function and the kaonic
wave functions. The B-meson wave functions $\Psi_B$ and $\bar\Psi_B$
up to next-to-leading order Fock state expansion depend on two
phenomenological parameters $\bar\Lambda$ and $\delta$. An estimate
of $\bar\Lambda$ using QCD sum rule approach gives
$\bar\Lambda=0.57\pm0.07 {\rm GeV}$ \cite{lambdavalue}. By comparing
the PQCD results of the $B\to\pi$ form factor with the QCD LCSR
results and the lattice QCD calculations, Ref.\cite{hwbpi} shows
that $\bar\Lambda=0.55\pm 0.05 {\rm GeV}$. As for the value of
$\delta$, it has been pointed out that if the contribution from the
B-meson three-particle wave function is limited to be within $\pm
20\%$ of that of the WW-like wave function within the energy region
of $Q^2\in [0,\sim 10 {\rm GeV}^2]$, then the value of $\delta$
should be restricted within the region of $[0.25,0.30]$ \cite{hqw}.
For clarity, we take the same regions as obtained from the $B\to
\pi$ case \cite{hwbpi,hqw} for both $\bar\Lambda$ and $\delta$, i.e.
$\bar\Lambda \in [0.50,0.60] {\rm GeV}$ and $\delta\in[0.25,0.30]$,
to study the form factors $F_+^{B\to K}(q^2)$ and $F_0^{B\to
K}(q^2)$ in the large and intermediate energy regions. Furthermore,
according to the discussion in the last section, the remaining
uncertainty of the kaonic twist-2 wave function $\Psi_K$ is caused
by the value of $a^K_1(1{\rm GeV})$, cf. Eq.(\ref{psika1}). There we
take $a^K_1(1{\rm GeV}) = 0.05\pm0.02$ \cite{lcsr1} to do our
discussion. As for the twist-3 wave function $\Psi_p$, we take two
groups of parameters as shown in Eqs.(\ref{group1},\ref{group2}) to
do the calculation.

Next, we compare the ${\cal O}(1/m^2_b)$ result of the form factor
with the leading order one that is of order ${\cal O}(1/m_b)$ and is
calculated by using the WW-like B-meson wave function, and also make
a comparison with the LCSR results of Ref.\cite{nlosum,sumrule} in
the large and intermediate energy regions. Through comparison,
preferable values for the undetermined parameters can be found. The
$B\to K$ transition form factors $F^{B\to K}_+(q^2)$ and $F^{B\to
K}_0(q^2)$ have been studied within the framework of QCD LCSR
\cite{sumrule}, especially at $q^2=0$, it shows
\begin{equation} \label{zeroff}
F^{B\to K}_{+,0}(0)=0.331\pm0.041+0.25[a^K_1(1{\rm GeV})-0.17] ,
\end{equation}
e.g. when $a^K_1(1{\rm GeV})=0.05$, $F^{B\to
K}_{+,0}(0)=0.301\pm0.041$. More generally, $F^{B\to K}_+(q^2)$ and
$F^{B\to K}_0(q^2)$ can be parameterized in the following form
\cite{sumrule}:
\begin{equation}\label{sumruleapp}
F^{B\to K}_{+,0}(q^2)=f^{as}(q^2)+a^K_1(\mu_0)
f^{a^K_1}(q^2)+a^K_2(\mu_0) f^{a^K_2}(q^2)+a^K_4(\mu_0)
f^{a^K_4}(q^2),
\end{equation}
where $f^{as}$ contains the contributions to the form factor from
the asymptotic DA and all higher-twist effects from three-particle
quark-quark-gluon matrix elements, $f^{a^K_1,a^K_2,a^K_4}$ contains
the contribution from the higher Gegenbauer term of DA that is
proportional to $a^K_1$, $a^K_2$ and $a^K_4$ respectively. Here the
factorization scale $\mu_{0}$ should be taken as $2.2{\rm GeV}$,
since the functions $f^{as,a^K_1,a^K_2,a^K_4}$ are determined with
$\mu_{0}=2.2{\rm GeV}$ \cite{sumrule}. The explicit expressions of
$f^{as,a^K_1,a^K_2,a^K_4}$ can be found in Table V and Table IX of
Ref.\cite{sumrule}. For the Gegenbauer moments $a^K_2(2.2{\rm GeV})$
and $a^K_4(2.2{\rm GeV})$, we take their preferred values:
$a^K_2(2.2{\rm GeV})=0.080$ and $a^K_4(2.2{\rm GeV})=-0.0089$
\cite{sumrule}. While for $a^K_1(2.2{\rm GeV})$, it equals to $0.793
a^K_1(1{\rm GeV})$ with the help of QCD evolution.

\subsection{Basic properties of the form factor up to ${\cal O}(1/m^2_b)$}

\begin{figure}
\centering
\includegraphics[width=0.5\textwidth]{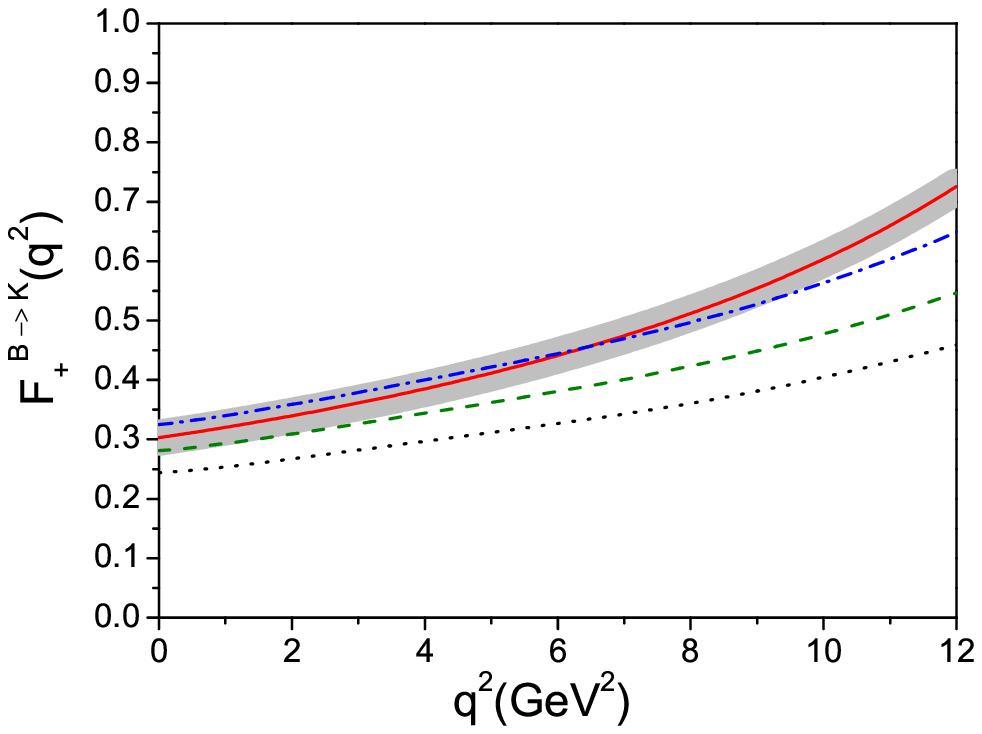}%
\includegraphics[width=0.5\textwidth]{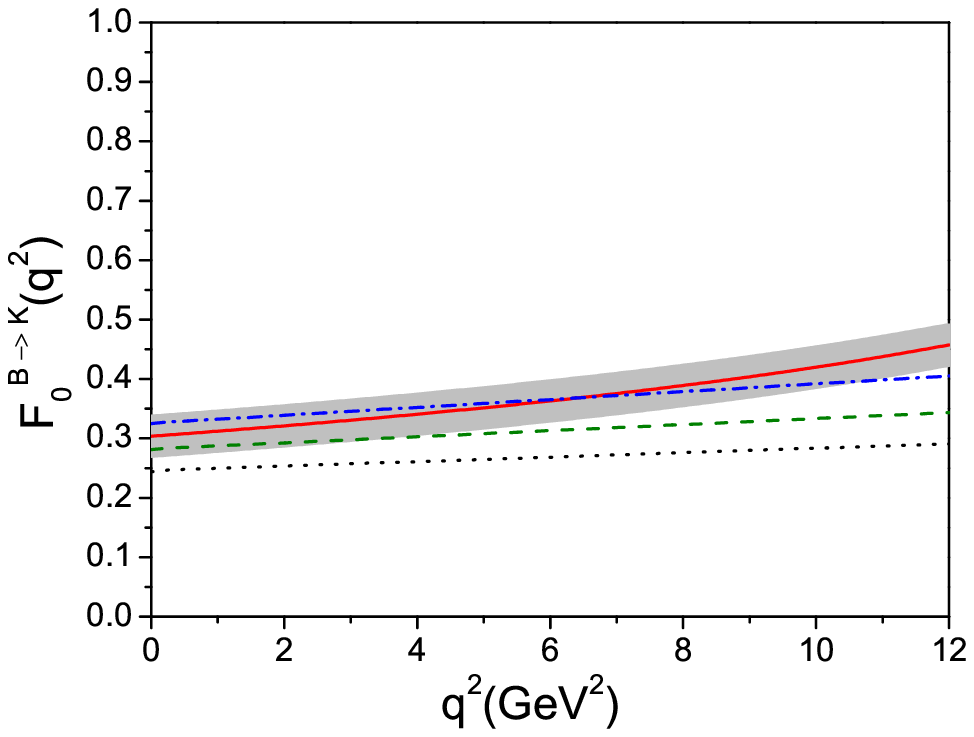}
\caption{PQCD results for the $B\to K$ transition form factors
$F_+^{B\to K}(q^2)$ (Left) and $F_0^{B\to K}(q^2)$ (Right) with
$\delta=0.275$ and $a^K_1(1{\rm GeV})=0.05$. The dash-dot line, the
dashed line and the dotted line stand for $\bar\Lambda=0.50$ GeV,
$0.55$ GeV and $0.60$ GeV respectively. For comparison, the solid
line comes from the QCD LCSR result as shown in
Eq.(\ref{sumruleapp}) and the fuscous shaded band shows its
theoretical error $\pm 10\%$.} \label{lambda}
\end{figure}

\begin{figure}
\centering
\includegraphics[width=0.5\textwidth]{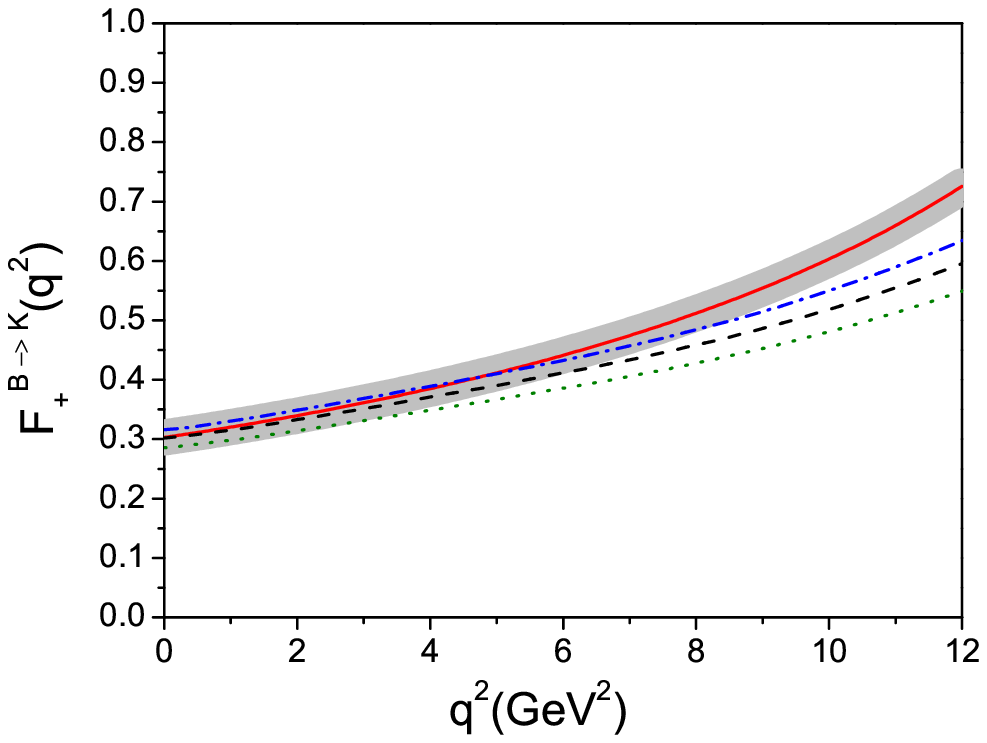}%
\includegraphics[width=0.5\textwidth]{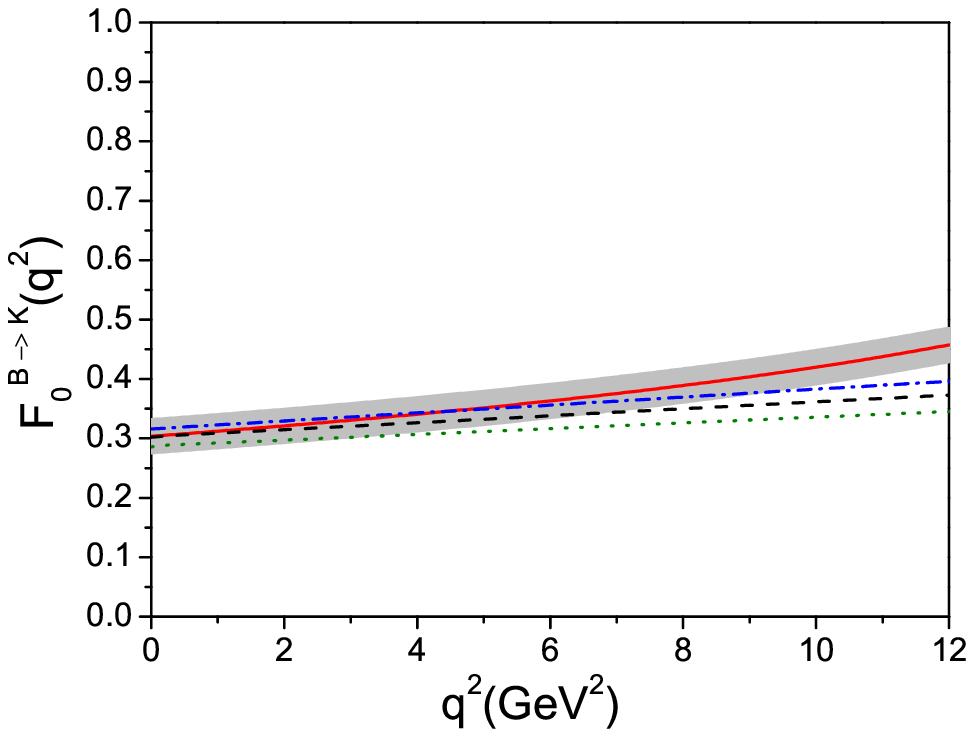}
\caption{PQCD results for the $B\to K$ transition form factors
$F_+^{B\to K}(q^2)$ (Left) and $F_0^{B\to K}(q^2)$ (Right) with
$\bar\Lambda=0.525$ GeV and $a^K_1(1{\rm GeV})=0.05$. The dotted
line, the dashed line and the dash-dot line stand for $\delta=0.25$,
$0.275$ and $0.30$ respectively. For comparison, the solid line
comes from the QCD LCSR as shown in Eq.(\ref{sumruleapp}) and the
fuscous shaded band shows its theoretical error $\pm 10\%$.}
\label{delta}
\end{figure}

First, we discuss the properties of $F_+^{B\to K}(q^2)$ and
$F_0^{B\to K}(q^2)$ caused by the B-meson wave function. For such
purpose, we fix the kaonic wave functions by setting $a^K_1(1{\rm
GeV})=0.05$ and by using the Group 1 parameters for $\Psi_p$. We
show the $B\to K$ transition form factors $F_+^{B\to K}(q^2)$ and
$F_0^{B\to K}(q^2)$ with $\delta=\delta_{c}=0.275$ in
Fig.(\ref{lambda}), where $\bar\Lambda$ varies within the region of
$[0.5{\rm GeV},0.6{\rm GeV}]$. For comparison, we show the QCD LCSR
result with $a^K_1(1{\rm GeV})=0.05$ and its theoretical error
($\sim\pm 10\%$) \cite{sumrule} by a fuscous shaded band in
Fig.(\ref{lambda}). The results show that the $B\to K$ transition
form factors will decrease with the increment of $\bar\Lambda$. And
the best fit of the QCD LCSR result at $q^2=0$ shows that
$\bar\Lambda\cong \bar\Lambda_{c}=0.525 {\rm GeV}$. Moreover, we
show $F_+^{B\to K}(q^2)$ and $F_0^{B\to K}(q^2)$ with
$\bar\Lambda=\bar\Lambda_{c}=0.525$ GeV in Fig.(\ref{delta}), where
$\delta$ varies within the region of $[0.25,0.30]$. The results show
that the $B\to K$ transition form factors will increase with the
increment of $\delta$. It can be found that when setting
$a^K_1(1{\rm GeV})=0.05$, and by varying $\delta$ within the region
of $[0.25,0.30]$ and $\bar\Lambda$ within the region of $[0.5{\rm
GeV},0.6{\rm GeV}]$, $F^{B\to K}_{+,0}(0)$ runs within the region of
$[0.23,0.34]$. since the best agreement between the PQCD result and
the QCD LCSR result at $q^2=0$ is obtained around
$\bar\Lambda_c=0.525 {\rm GeV}$ and $\delta_c=0.275$, we shall
always take $\bar\Lambda=\bar\Lambda_c$ and $\delta=\delta_c$ to do
our following calculations if not specially stated.

\begin{figure}
\centering
\includegraphics[width=0.5\textwidth]{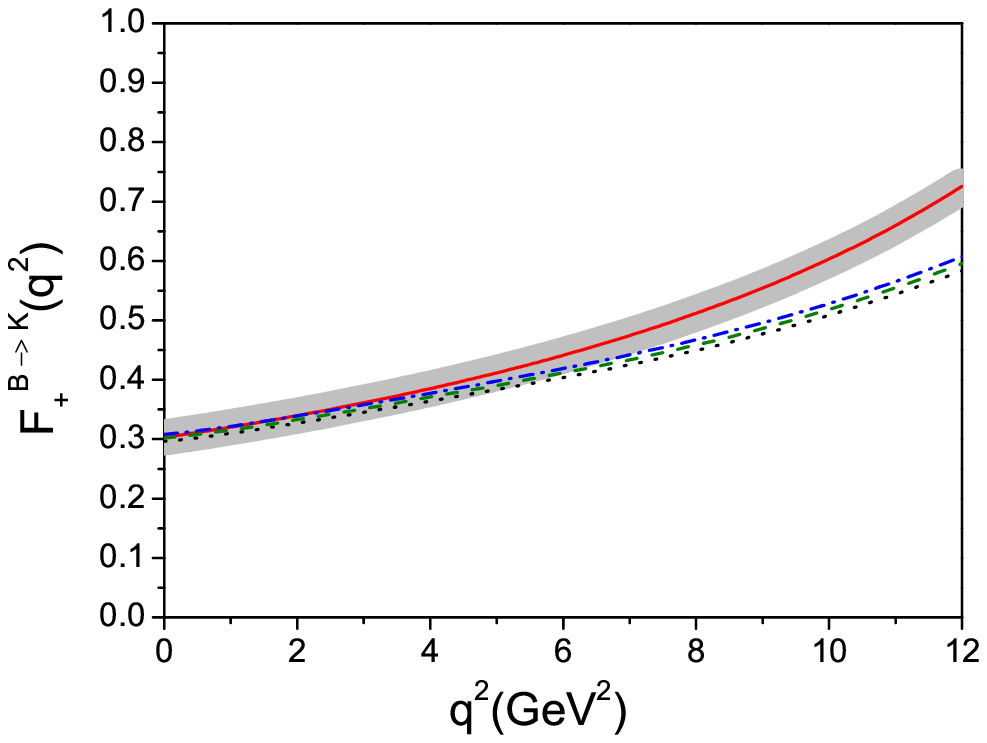}%
\includegraphics[width=0.5\textwidth]{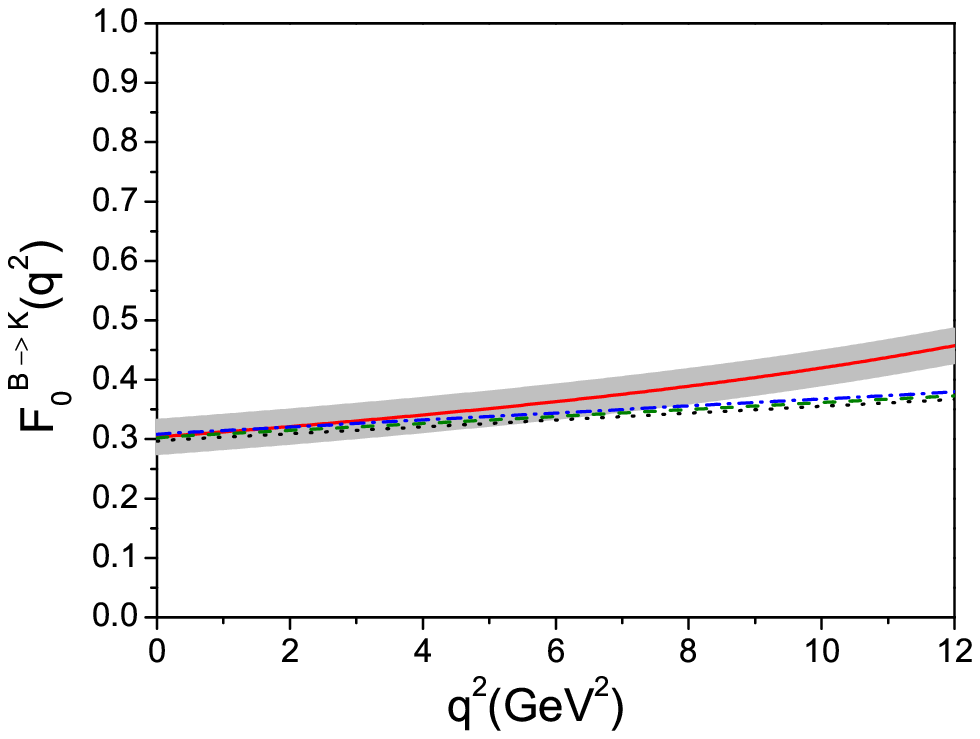}
\caption{PQCD results for the $B\to K$ transition form factors
$F_+^{B\to K}(q^2)$ (Left) and $F_0^{B\to K}(q^2)$ (Right) with
$\bar\Lambda=0.525$ GeV and $\delta=0.275$. The dotted line, the
dashed line and the dash-dot line stand for $a^K_1(1{\rm
GeV})=0.03$, $0.05$ and $0.07$ respectively.} \label{a1k}
\end{figure}

\begin{figure}
\centering
\includegraphics[width=2.9in]{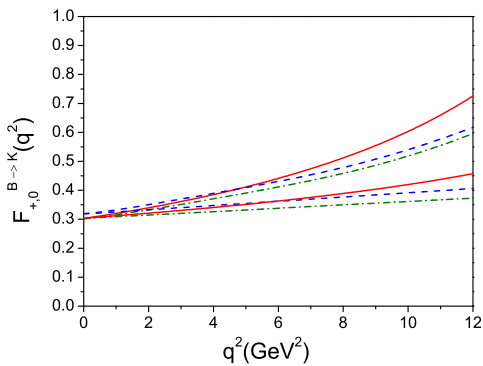}
\caption{PQCD results for the $B\to K$ transition form factors
$F_+^{B\to K}(q^2)$ and $F_0^{B\to K}(q^2)$ with
$\bar\Lambda=0.525GeV$, $\delta=0.275$ and $a^K_1(1{\rm GeV})=0.05$.
The dash-dot and the dashed lines are for $\Psi_p$ with Group 1
parameters Eq.(\ref{group1}), Group 2 parameters Eq.(\ref{group2})
respectively. For comparison, the solid lines come from the QCD LCSR
with $a^K_1(1{\rm GeV})=0.05$ \cite{sumrule}. } \label{twistp}
\end{figure}

Second, we discuss the properties of $F_+^{B\to K}(q^2)$ and
$F_0^{B\to K}(q^2)$ caused by the twist-2 wave function $\Psi_K$,
i.e. by the value of $a^K_1(1{\rm GeV})$. For such purpose, we fix
the B-meson wave functions by setting $\delta=\delta_c$ and
$\bar\Lambda=\bar\Lambda_c$ and by using the Group 1 parameters for
$\Psi_p$. We show the $B\to K$ transition form factors $F_+^{B\to
K}(q^2)$ and $F_0^{B\to K}(q^2)$ in Fig.(\ref{a1k}) with
$a^K_1(1{\rm GeV})=0.03$, $0.05$ and $0.07$ respectively. It can be
found that the form factors shall be increased with the increment of
$a^K_1(1{\rm GeV})$, which agree with the observation of
Ref.\cite{sumrule}. Furthermore, since the contribution from
$\Psi_p$ is sizable to that of $\Psi_K$, it is necessary to make a
discussion on its uncertainty to the $B\to K$ transition form
factor. Fig.(\ref{twistp}) shows $F_+^{B\to K}(q^2)$ and $F_0^{B\to
K}(q^2)$ with two groups of parameters for $\Psi_p$. The results are
very close to each other due to the close shape of their $\phi_p$ as
shown in Fig.(\ref{DAphip}), e.g. around the region of $q^2\sim 0$
the difference between them is less than $6\%$. So by taking proper
transverse momentum dependence for the wave function $\Psi_p$, where
we have taken the BHL prescription for its transverse momentum
dependence, the uncertainties from its distribution amplitude
$\phi_p$ can be reduced.

\begin{figure}
\centering
\begin{minipage}[c]{0.48\textwidth}
\centering
\includegraphics[width=2.9in]{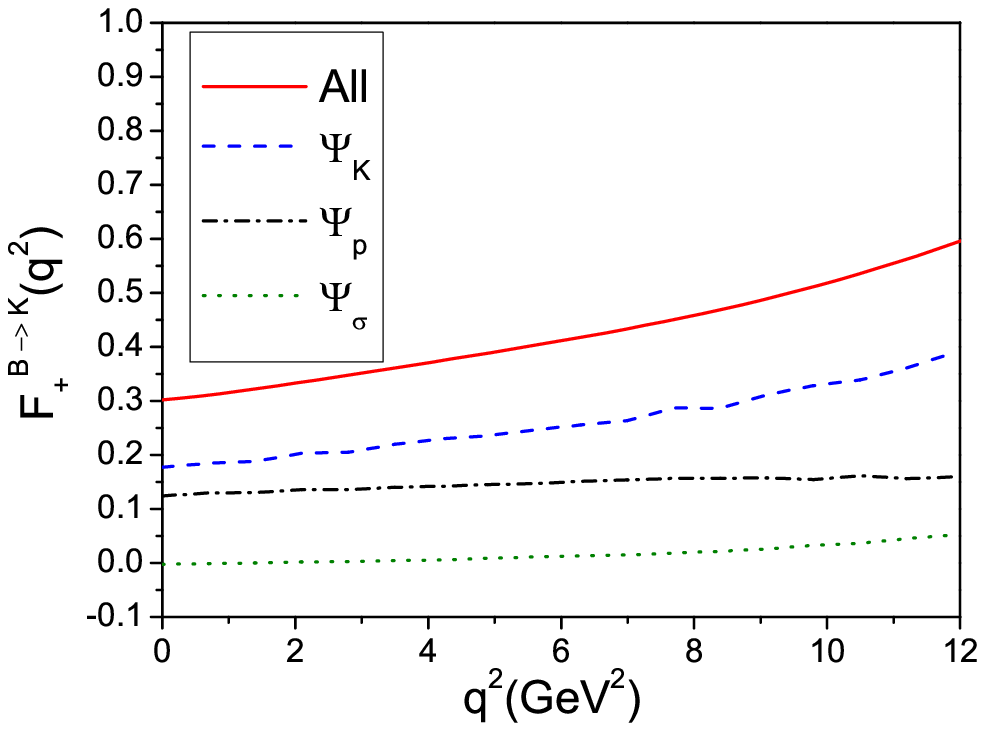}
(a)
\end{minipage}%
%\hspace{0.1in}
\begin{minipage}[c]{0.48\textwidth}
\centering
\includegraphics[width=2.9in]{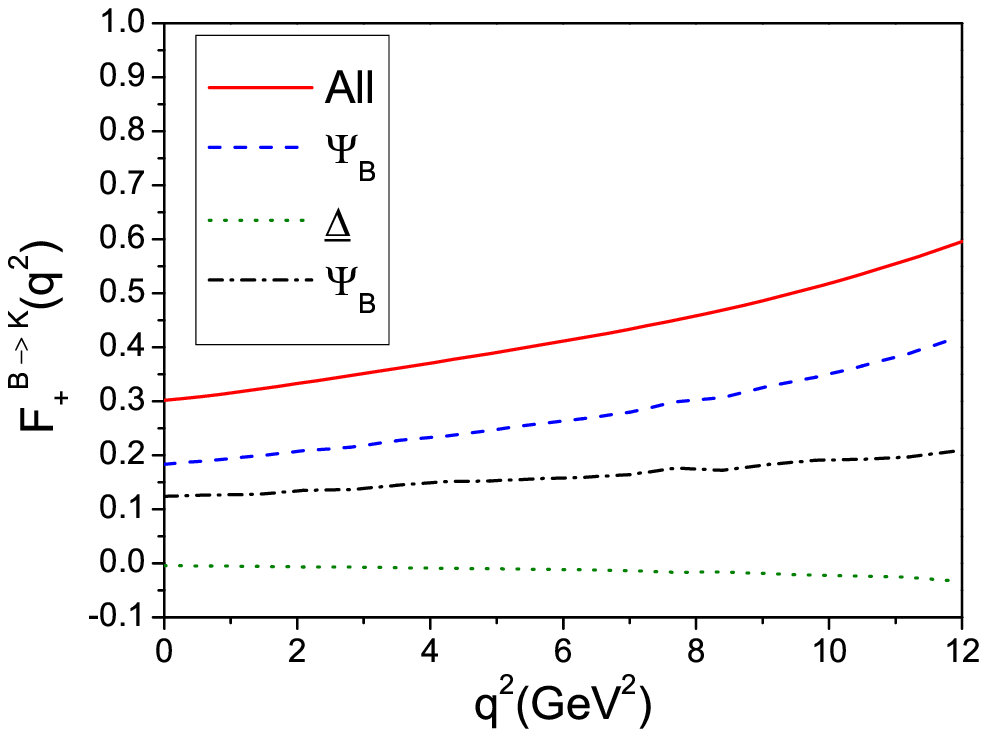}
(b)
\end{minipage}
\caption{PQCD results for the $B\to K$ transition form factor
$F_+^{B\to K}(q^2)$ with fixed $\bar\Lambda=0.55GeV$, $\delta=0.275$
and $a^K_1(1{\rm GeV})=0.05$. The left diagram is for the different
kaon twist structures, $\Psi_K$, $\Psi_p$ and $\Psi_\sigma$. The
right diagram is for the different B meson structures, $\Psi_B$,
$\bar\Psi_B$ and $\Delta$.} \label{kaonwave}
\end{figure}

Finally, in order to get a deep understanding of the $B\to K$
transition form factor, we discuss the contributions from different
parts of the B-meson wave function or the kaon wave function,
correspondingly. Here we take $F_+^{B\to K}(q^2)$ to do our
discussions and the case of $F_0^{B\to K}(q^2)$ can be done in a
similar way. For convenience, we set $\bar\Lambda=\bar\Lambda_c$,
$\delta=\delta_c$, $a^K_1(1{\rm GeV})=0.05$ and by using the Group 1
parameters for $\Psi_p$. When discussing the contribution from one
of the kaon wave function structures, the contribution from all the
B-meson wave function structures are summed up, and vice versa.
Fig.(\ref{kaonwave}a) shows the contributions from the different
twist structures of the kaon wave function, i.e. $\Psi_K$, $\Psi_p$
and $\Psi_\sigma$ (the contributions from the terms involving
$\Psi'_\sigma$ are included in $\Psi_\sigma$), respectively. One may
observe that the contribution from $\Psi_p$ is comparable to that of
$\Psi_K$, e.g. its contribution is about $70\%$ of that of $\Psi_K$
at $q^2\simeq 0$, and the contribution from $\Psi_\sigma$ is small.
Fig.(\ref{kaonwave}b) presents the contributions from $\Psi_B$,
$\bar\Psi_B$ and $\Delta$ respectively. It can be found that the
contribution from $\bar\Psi_B$ is about $50\%-67\%$ in comparison to
that of $\Psi_B$ in the region of $q^2\in [0,10 {\rm GeV}^2]$, while
the contribution from $\Delta$ is negligible in comparison to that
of $\Psi_B$ and $\bar\Psi_B$. So the contribution from $\bar\Psi_B$
should be included for a consistent estimation to the next leading
order. As a comparison, it can be found that under the leading order
estimation the contribution from $\bar\Psi_B$ is only about $20\%$
in comparison to that of $\Psi_B$ at $q^2=0$, which agree with the
rough order estimation that the contribution from $\bar\Psi_B$ is of
order ${\cal O}(1/m_b)$. So as to the leading order estimation
${\cal O}(1/m_b)$, the contribution from $\bar\Psi_B$ is usually
neglected in the literature. Such difference of $\bar\Psi_B$'s
contribution between the leading order estimation and the
next-to-leading order estimation is mainly due to the fact that the
transverse momentum dependence of the B-meson wave functions are
merely a delta function under the WW-approximation (the
leading-order estimation), while it shall be broadened to a certain
degree according to the value of $\delta$ by taking into account the
3-particle Fock states' contributions (the next-to-leading order
estimation), cf. fig.(2) of Ref.\cite{hqw}. So qualitatively, the
contributions from $\bar{\Psi}_B$ shall be raised to a certain
degree for the next-to-leading order case, due to the less
suppression of the end-point region ($\xi\to 0$) from the transverse
momentum distributions than that of the leading order case. And then
the naive order estimation for the contribution of $\bar\Psi_B$ is
no longer correct, and the contributions from $\Psi_B$ and
$\bar\Psi_B$ are both important in the next-to-leading order
calculation.

\subsection{Comparison with the leading order results}

The WW-like B-meson wave functions in the compact parameter
$b_B$-space can be found in Ref.\cite{hwbpi}. Taking the WW-like
wave functions and cutting off the power suppressed terms in the
hard scattering amplitude, we can obtain the leading order results
(${\cal O}(1/m_b)$) for the form factors $F_+^{B\to K}(q^2)$ and
$F_0^{B\to K}(q^2)$. Strictly, one should cut off the contribution
from $\bar\Psi_B$ to obtain the leading order estimation, since
$\bar\Psi_B$ is power suppressed in comparison to $\Psi_B$. However
for easy comparison with the results in the literature, e.g.
Ref.\cite{lucai}, we keep $\bar\Psi_B$ in the leading order
estimation. For convenience, we take $\bar\Lambda=\bar\Lambda_c$,
$\delta=\delta_c$, $a^K_1(1{\rm GeV})=0.05$ and by using the Group 1
parameters for the wave function $\Psi_p$ to do a comparison of the
leading order results with the total results that include the
contributions up to order ${\cal O}(1/m^2_b)$. It can be found that
the leading order results are smaller than the total results by
about $25\%$ in the large recoil region, e.g. at $q^2=0$, the
leading order $F_{+,0}^{B\to K}(0)=0.229$. One may observe that a
larger leading order estimation has been obtained in
Ref.\cite{lucai}, which shows $F_{+,0}^{B\to K}(0)=0.321\pm0.036$.
We argue that the present leading order estimation is more reliable,
and the larger value of $F_{+,0}^{B\to K}(0)$ derived in
Ref.\cite{lucai} is mainly due to the following two reasons: 1) Even
though the Sudakov and threshold resummation factors shall kill the
endpoint singularity of the process \cite{lis,lucai,li1,charng}, the
transverse momentum dependence of kaonic wave functions are still
important to give a more reliable PQCD estimation, which is similar
to the cases of $B\to \pi$ form factor \cite{hwbpi} and the pion
electromagnetic form factor \cite{pitwist3}. In Ref.\cite{lucai} the
transverse momentum dependence of kaonic wave functions are lacking,
i.e. the distribution amplitude other than the wave function is
used. While in our present calculation, the BHL-prescription is
adopted for the kaonic transverse momentum dependence. As for the
wave function $\Psi_p(x,\mathbf{k}_\perp)$, such transverse momentum
dependence will results in a double humped DA $\phi_p$ as shown in
Fig.(\ref{DAphip}) and then it shall give more effective suppression
in the end-point region than the one used in Ref.\cite{lucai}. In
fact, it can be found that the contributions from the end point
region shall always be overestimated without taking the transverse
momentum into the twist-3 wave function $\Psi_p(x,\mathbf{k}_\perp)$
\footnote{For example, a detailed discussion on the model dependence
of pionic twist-3 wave function $\Psi_p(x,\mathbf{k}_\perp)$ can be
found in Ref.\cite{pitwist3}.}. Furthermore, by taking proper
transverse momentum dependence for the wave function $\Psi_p$, the
uncertainties from its distribution amplitude $\phi_p$ can be
reduced as has been discussed in Sec.III.A; 2) the distribution
amplitude of $\Psi_K$ with a much bigger value of $a^K_1(1{\rm
GeV})$, i.e. $a^K_1(1{\rm GeV})=0.17$, is adopted by
Ref.\cite{lucai}. Since the form factors increases with the
increment of $a^K_1(1{\rm GeV})$, a larger value of $a^K_1(1{\rm
GeV})$ shall increase the form factors.

Furthermore, by varying $\bar\Lambda$ within the region of
$[0.50,0.55]$, the uncertainty caused by $\bar\Lambda$ is the
biggest and is of order $(1/m_b)$. While by varying $\delta$ within
the region of $[0.25,0.30]$, the uncertainty caused by $\delta$ is
smaller and are of order $(1/m_b^2)$. This can be qualitatively
explained as that $\bar\Lambda$ is the characteristic parameter that
determines the leading Fock-state behavior of the $B$-meson wave
functions, while $\delta$ is the characteristic parameter that
determines the higher Fock-state's behavior of the $B$-meson wave
functions. The uncertainties from $a^K_1$ and $\Psi_K$ are less than
$10\%$ in the large recoil region.

\subsection{Comparison with the LCSR results}

The $B\to K$ transition form factor have been analyzed by several
groups under the QCD LCSR approach
\cite{huangli,sumrule,pballsum1,nlosum}. New sum rule for $B\to K$
is derived from the correlation functions expanded near the light
cone in terms of B-meson distributions \cite{nlosum}, in which the
contributions of the quark-antiquark and quark-antiquark-gluon
components in the B-meson are taken into account. It has been found
that the $B\to K$ transition form factor in the large recoil region
does not receive contributions from the 3-particle B-meson DA's. One
may observe that if substituting the B-meson DAs, which are derived
by doing the integration over $b_B$ in
Eqs.(\ref{newmodel1},\ref{newmodel2}), into the formulae of
Ref.\cite{nlosum}, then one can obtain the same results as that of
Ref.\cite{nlosum}, since our B-meson DAs are close to the
exponential model wave functions adopted in Ref.\cite{nlosum}.
Furthermore, one may observe that the result of $F^{B\to
K}_{+,0}(0)=0.31\pm 0.04$ under the condition of $a^K_1(1{\rm
GeV})=0.05\pm0.03$ agrees well with our present PQCD estimation.
Secondly, a systematic QCD LCSR calculation has been done in
Ref.\cite{sumrule} by including the one-loop radiative corrections
to the twist-2 and twist-3 contributions, and leading-order twist-4
corrections. Some comparison of their results with our present one
can be found in Figs.(\ref{lambda},\ref{delta},\ref{a1k}), which
also shows a good agreement within reasonable errors. For example,
from Eq.(\ref{sumruleapp}) it can be found that the uncertainty of
form factor caused by $a^K_1(1{\rm GeV})$ within the region of
$[0.03,0.07]$ is less than $5\%$, which is consistent with our
present result as shown in Fig.(\ref{a1k}).

\section{discussion and summary}

In this paper, we have examined the $B\to K$ transition form factor
in the PQCD approach up to order ${\cal O}(1/m^2_b)$, where the
transverse momentum dependence for the wave function, the Sudakov
effects and the threshold effects are included to regulate the
endpoint singularity and to derive a more reasonable result. We have
confirmed that the PQCD approach can be applied to calculate the
$B\to K$ transition form factor in the large recoil regions. We
emphasize that the transverse momentum dependence for both the B
meson and the kaon is important to give a better understanding of
the $B\to K$ transition form factor. Fig.(\ref{kaonwave}a) shows
that the contribution from the pionic twist-3 wave function $\Psi_p$
is sizable in comparison to that of $\Psi_K$, and the contribution
from $\Psi_\sigma$ is small. While Fig.(\ref{kaonwave}b) shows that
by using the B-meson wave functions up to next-to-leading order Fock
state expansion, the contributions from $\Psi_B$ and $\bar\Psi_B$
are important.

In Refs.\cite{hwbpi,hqw}, we have shown that the results from the
PQCD approach, the lattice QCD approach and the QCD LCSRs are
complementary to each other and by combining the results of those
three approaches, one can obtain an understanding of the $B\to\pi$
transition form factor in the whole physical regions. And the best
fit of the PQCD results with that of the QCD LCSR results in the
large recoil region can be obtained by taking
$\bar\Lambda\in[0.50,0.60]$ and $\delta\in[0.25,0.30]$ \cite{hqw}.
In the present paper, we show that within the regions of
$\bar\Lambda\in[0.50,0.55]$, $\delta\in[0.25,0.30]$ and $a^K_1(1{\rm
GeV})\in[0.03,0.07]$, the PQCD results on the $B\to K$ form factor
in the large recoil region also agree well with that of the QCD LCSR
results \cite{sumrule,nlosum}. Our present PQCD results in some
sense is more reliable than the LCSR calculations due to the fact
that by taking the transverse momentum dependence properly for the
wave functions the soft endpoint singularity have been effectively
suppressed, e.g. as is shown in Sec.III.A the difference between the
two models for $\Psi_p$ is less than $6\%$ in the large recoil
region, while for the LCSR approach large uncertainty comes from the
kaonic twist-3 DA $\phi_p$ that is not too well-known. By running
the parameters within the above regions, we obtain $F^{B\to
K}_{+,0}(0)=0.30\pm0.04$. Finally, to illustrate the
$SU_f(3)$-breaking effects, we calculated the ratio with the help of
the $B\to \pi$ results in Ref.\cite{hqw}: $[F^{B\to
K}_{+,0}(0)/F^{B\to \pi}_{+,0}(0)]=1.13\pm0.02$, which favors a
small $SU_f(3)$-breaking effects.

\begin{center}
\section*{Acknowledgements}
\end{center}

This work was supported in part by the Natural Science Foundation of
China (NSFC) and by the Grant from Chongqing University. This work
was also partly supported by the Natural Science Foundation of
Chongqing under the Grant NO. 8562 and the National Basic Research
Programme of China under Grant NO. 2003CB716300. We thank Prof.C.D.
Lu for helpful comments. \\

\end{document}